\documentclass[a4paper,12pt,fleqn]{article}
\usepackage{changepage}
\usepackage{amssymb}
\usepackage{amsmath}
\usepackage[margin=1.0in]{geometry}
\usepackage{graphicx}
\usepackage{dcolumn}
\usepackage{bm}
\usepackage{hyperref}
\usepackage{textgreek}

\begin{document}

\newcommand\p{\partial}
\newcommand\Psibar{\overline{\Psi}}
\newcommand\psibar{\overline{\psi}}
\newcommand\chibar{\overline{\chi}}
\newcommand\Ebar{\overline{E}}
\newcommand\Fbar{\overline{F}}
\newcommand\Mbar{\overline{M}}
\newcommand\Qbar{\overline{Q}}
\newcommand\Tbar{\overline{T}}
\newcommand\Vbar{\overline{V}}
\newcommand\xbar{\overline{x}}
\newcommand\jbar{\overline{j}}
\newcommand\kbar{\overline{k}}
\newcommand\ybar{\overline{y}}
\newcommand\lambar{\overline{\lambda}}
\newcommand\sigbar{\overline{\sigma}}
\newcommand\omegbar{\overline{\omega}}
\newcommand\rmi{\mathrm{i}}
\newcommand\rmc{\mathrm{c}}
\newcommand\rme{\mathrm{e}}
\newcommand\rmexp{\mathrm{exp}}
\newcommand\rmln{\mathrm{ln}}
\newcommand\rmsin{\mathrm{sin}}
\newcommand\rmcos{\mathrm{cos}}
\newcommand\rmsinh{\mathrm{sinh}}
\newcommand\rmcosh{\mathrm{cosh}}
\newcommand\rmtan{\mathrm{tan}}
\newcommand\rmarcsin{\mathrm{arcsin}}
\newcommand\rmarccos{\mathrm{arccos}}
\newcommand\rmarctan{\mathrm{arctan}}
\newcommand\rmlim{\mathrm{lim}}
\newcommand\rmTr{\mathrm{Tr}}
\newcommand\rmdet{\mathrm{det}}
\newcommand\rmdiag{\mathrm{diag}}
\newcommand\rmker{\mathrm{ker}}
\newcommand\rmsgn{\mathrm{sgn}}
\newcommand\rmint{\mathrm{int}}
\newcommand\rmMD{\mathrm{MD}}
\newcommand\rmic{\mathrm{ic}}
\newcommand\rmt{\mathrm{t}}
\newcommand\rmT{\mathrm{T}}
\newcommand\rmd{\mathrm{d}}
\newcommand\rmC{\mathrm{C}}
\newcommand\rmD{\mathrm{D}}
\newcommand\rmM{\mathrm{M}}
\newcommand\rmP{\mathrm{P}}
\newcommand\rmQ{\mathrm{Q}}
\newcommand\rmY{\mathrm{Y}}
\newcommand\rmconst{\mathrm{const.}}
\newcommand\rmem{\mathrm{em}}
\newcommand\rmRe{\mathrm{Re}}
\newcommand\rmIm{\mathrm{Im}}
\newcommand\scL{\mathcal{L}}
\newcommand\scF{\mathcal{F}}
\newcommand\sL{\mathscr{L}}
\newcommand\bux{\underline{\boldsymbol{x}}}
\newcommand\bua{\underline{\boldsymbol{a}}}
\newcommand\buu{\underline{\boldsymbol{u}}}
\newcommand\tV{\tilde{V}}
\newcommand\tS{\widetilde{S}}
\newcommand\tT{\widetilde{T}}
\newcommand\tW{\tilde{W}}
\newcommand\tP{\widetilde{P}}
\newcommand\tj{\tilde{j}}
\newcommand\ba{\boldsymbol{a}}
\newcommand\bp{\boldsymbol{p}}
\newcommand\bE{\boldsymbol{E}}
\newcommand\bM{\boldsymbol{M}}
\newcommand\bV{\boldsymbol{V}}
\newcommand\bW{\boldsymbol{W}}
\newcommand\btau{\boldsymbol{\tau}}
\newcommand\balpha{\boldsymbol{\alpha}}
\newcommand\bPhi{\boldsymbol{\Phi}}
\newcommand\bT{\boldsymbol{T}}
\newcommand\br{\boldsymbol{r}}
\newcommand\be{\boldsymbol{e}}
\newcommand\bj{\boldsymbol{j}}
\newcommand\bk{\boldsymbol{k}}
\newcommand\boldm{\boldsymbol{m}}
\newcommand\bn{\boldsymbol{n}}
\newcommand\bt{\boldsymbol{t}}
\newcommand\bv{\boldsymbol{v}}
\newcommand\bw{\boldsymbol{w}}
\newcommand\bep{\boldsymbol{\epsilon}}
\newcommand\bphi{\boldsymbol{\phi}}
\newcommand\xhat{\hat{\boldsymbol{x}}}
\newcommand\yhat{\hat{\boldsymbol{y}}}
\newcommand\zhat{\hat{\boldsymbol{z}}}
\newcommand\rhat{\hat{\boldsymbol{r}}}
\newcommand\sdual{{}^{*}\!s}
\newcommand\Sdual{{}^{*}\!S}
\newcommand\spart{\slashed{\partial}}
\newcommand\lspart{\overleftarrow{\slashed{\partial}}}

\section*{{\LARGE The self-coupled Einstein-Cartan-Dirac equations in terms of Dirac bilinears}}
\vspace{0.5cm}



\begin{adjustwidth}{1cm}{}
{\large \textbf{S M Inglis and P D Jarvis}}

\vspace{0.2cm}
{\small\noindent School of Natural Sciences, University of Tasmania, Private Bag 37, Hobart, Tasmania, 7001, Australia}

\vspace{0.2cm}
{\small\noindent Email: {\tt Shaun.Inglis@utas.edu.au, Peter.Jarvis@utas.edu.au}}

\vspace{0.5cm}
{\small
\noindent\textbf{Abstract.} In this article we present the algebraic rearrangement, or matrix inversion of the Dirac equation in a curved Riemann-Cartan spacetime with torsion; the presence of non-vanishing torsion is implied by the intrinsic spin-1/2 of the Dirac field. We then demonstrate how the inversion leads to a reformulation of the fully non-linear and self-interactive Einstein-Cartan-Dirac field equations in terms of Dirac bilinears. It has been known for some decades that the Dirac equation for charged fermions interacting with an electromagnetic field can be algebraically inverted, so as to obtain an explicit rational expression of the four-vector potential of the gauge field in terms of the spinors. Substitution of this expression into Maxwell's equations yields the bilinear form of the self-interactive Maxwell-Dirac equations. In the present (purely gravitational) case, the inversion process yields \emph{two} rational four-vector expressions in terms of Dirac bilinears, which act as gravitational analogues of the electromagnetic vector potential. These ``potentials'' also appear as irreducible summand components of the connection, along with a traceless residual term of mixed symmetry. When taking the torsion field equation into account, the residual term can be written as a function of the object of anholonomity. Using the local tetrad frame associated with observers co-moving with the Dirac matter, a generic vierbein frame can described in terms of four Dirac bilinear vector fields, normalized by a scalar and pseudoscalar field. A corollary of this is that in regions where the Dirac field is non-vanishing, the self-coupled Einstein-Cartan-Dirac equations can in principle be expressed in terms of Dirac bilinears only.}
\end{adjustwidth}


                             

\section{\label{Introduction}Introduction}
The Dirac equation, the relativistic wave equation for spin-1/2 fermions, can be made to describe particles interacting with a gauge field by replacing the partial derivative with the covariant derivative for the particular field. For a gauge potential of a given form, the Dirac equation may be solved for the spinor field corresponding to the fermion state. One example solution for an electron in an external field is that of the hydrogen atom, where the Dirac equation correctly predicts fine structure as a result of relativistic corrections to the Hamiltonian \cite{LandauLifshitz1971}. However, the external Dirac-Coulomb solution itself does not explain the famous Lamb shift, which requires a consideration of how radiative corrections provided by the Maxwell field affect the energy of the bound electron \cite{Bethe1947}.

An inversion of the Dirac equation can be performed via algebraic rearrangement, such that the gauge potential is written as a rational, explicit function of the spinors \cite{Eliezer1958}. The outcome of this rearrangement procedure appears as though we have performed a matrix inversion, since the Dirac equation can be written in the form
\begin{equation}\label{Dirac electromagnetic matrix equation}
MA=R,
\end{equation}
where the complex $4\times4$ spinor-vector matrix $M$ is a function only of the components of the Dirac spinor. Assuming the vector potential $A$ is real, $M$ is invertible, and an explicit expression for $A$ in terms of the spinors can be obtained \cite{BoothLeggJarvis2001}. Substituting the inverted Dirac equation into the equations of motion for the gauge field results in a self-coupled system, where the charged fermion field interacts with itself in an internally consistent way. A central aspect of the algebraic inversion procedure is that the spinors do not appear as stand-alone objects, but rather as \emph{bilinear} combinations. An early proponent of using the bilinear description of Dirac states as the objects of primary interest was Takabayasi \cite{Takabayasi1957}, who promoted the idea of a relativistic hydrodynamical model of Dirac matter. The states of this model were not spinors or wavefunctions, but \emph{tensors} corresponding to quantum observables, such as current and spin densities. This in effect was an early substantial attempt to formulate a \emph{semi-classical} fluid model of relativistic quantum electrodynamics.

There exists a rich set of interrelationships between quadratic combinations of Dirac bilinears, known as \emph{Fierz identities} \cite{Fierz1937}, \cite{Lounesto2001} (alternatively, Fierz-Pauli-Kofink identities \cite{Baylis1996}); derived via a successive set of \emph{Fierz expansions} over a Dirac Clifford algebra primitive set of sixteen basis elements. Using a similar process, Crawford showed that \cite{Crawford1985}, given a set of sixteen bilinears formed from this set, the spinor field is recoverable up to a constant spinor with arbitrary phase. In addition to this set, there are two bilinears which are the real and imaginary parts of a complex bilinear (constructed with a both charge conjugated and a regular spinor: $\psibar{}^{\rmc}\gamma^{a}\psi$), and comprise a locally orthonormal tetrad frame along with the standard Lorentz four-vector and axial-vector fields \cite{Takahashi1983}.


In the electromagnetic case, the self-coupled \emph{Maxwell-Dirac} equations were shown to be describable in terms of the gauge \emph{independent} bilinears only, by Inglis and Jarvis \cite{InglisJarvis2014}, manifestly reflecting the physical gauge invariance of the system. Furthermore, these equations were able to be greatly simplified via the applications of infinitesimal invariance under several subgroups of the Poincar\'{e} group. These subgroups were chosen from a set of 158 given by Patera, Winternitz \& Zassenhaus \cite{PateraWinternitzZassenhaus1975}, where a comprehensive list of all the Poincar\'{e} Lie subalgebras and their corresponding generators are given. These symmetry reductions aid in the search for solutions to an otherwise intractable set of non-linear equations.

The ability to invert the Dirac equation is not limited to the electromagnetic case either, and we showed in a previous publication \cite{InglisJarvis2012} that an inversion can be performed for the non-Abelian gauge field $SU(2)$. We found that the algebraic process was very similar to the Abelian case, but with some extra difficulty, and the inverted form was given implicitly. It is currently unclear whether a similar generalisation exists for the strong $SU(3)$ case, although the $SU_{L}(2)\times U(1)$ electroweak case appears to be promising. Substitution into the Yang-Mills equations yields a fully non-linear self-interactive non-Abelian hydrodynamical theory, relevant to the study of non-perturbative high-energy plasmas. Another, simpler approach to modelling aspects of non-Abelian hydrodynamics, is to generalize the classical fluid mechanical equations to include local internal symmetries. A description of the non-Abelian Lorentz force involving chromoelectric and chromomagnetic field couplings was obtained this way in \cite{BistrovicJackiwNairPi2003}.

In this paper, we demonstrate how the Dirac equation in a curved Riemann-Cartan spacetime with torsion can be algebraically inverted, in an analogous manner to the $U(1)$ and $SU(2)$ cases; the covariant derivative we use contains the connection contracted with the generator for Lorentz transformations. Due to the intrinsic spin carried by the Dirac fermions, we consider the torsion field generated from the spin current density to be non-vanishing in general \cite{BlagojevicHehl2013}; an extra set of constraints on the gravitational field are obtained as a result. In comparison with (\ref{Dirac electromagnetic matrix equation}), the curved spacetime Dirac equation is of the form
\begin{equation}\label{Dirac gravitational matrix equation}
M\Omega+N\Omega_{5}=R,
\end{equation}
and the matrix inversion procedure yields explicit rational expressions for the gravitational ``vector potentials'' $\Omega$ and $\Omega_{5}$. It is important to note that unlike in the $U(1)$ electromagnetic case (\ref{Dirac electromagnetic matrix equation}), these $\Omega$-potentials are not exactly the gauge fields of the gravitational field. In fact, the inversions we obtain act to constrain two out of three irreducible components of the \emph{connection}, the local Lorentz gauge field. Both the connection and vierbein (corresponding to local translations) are the gauge fields of the Poincar\'{e} gauge theory of gravity \cite{BlagojevicHehl2013}. The study of Dirac equation inversion-based constraints for cases beyond the basic Einstein-Cartan system studied here are topics for future consideration.

In section \ref{The Einstein-Cartan-Dirac equations and conventions}, we derive an equation of the form (\ref{Dirac gravitational matrix equation}) from the standard Dirac equation in curved spacetime. We do this by considering an irreducible decomposition of the connection in $GL(4)$, which can be written as a sum of three terms. The trace term is a function of $\Omega_{a}$, and the two traceless terms are a fully antisymmetric function of $\Omega_{5a}$ and a residual term of mixed symmetry, ${}^{(3)}\Gamma_{abc}$.


In section \ref{The inversion procedure}, we give our definition of the tensor fields resulting from sandwiching elements of the Dirac Clifford algebra basis between Dirac spinors. Using this notation, we then show that by left-multiplying the curved spacetime Dirac equation and its charge conjugate with four different spinors, the resulting set of four equations can be solved explicitly for the \emph{two} gravitational vector potentials. These expressions are rational functions of bilinears and their first derivatives, but are not able to be expressed in terms of our tensor field set without further calculations.

The process by which we can write the inverted expressions in terms of bilinear tensor fields is given in \ref{Bilinear refinement using Fierz identities}. Here, we give a brief outline of the process by which Fierz expansions, where an outer product of two Dirac spinors is expanded in the Dirac Clifford algebra basis, are used to derive Fierz identities which are quadratic in the bilinears. These identities are then used to eliminate the explicit appearance of Dirac spinors in the inverted forms of the Dirac equation, replacing them with pure tensor expressions.

Section \ref{The Einstein-Cartan-Dirac self-coupled system} is given in four parts. In the first two parts, we describe the field equations of the Einstein-Cartan system, for the gravitational dynamics of space-time curvature and spin-torsion respectively. Expressions for the Ricci tensor, scalar, and the torsion are given in terms of the connection and the object of anholonomity. In the third part, we show how the algebraic torsion field equation can be used to place constraints on $\Omega_{a}$ and $\Omega_{5a}$, and to derive an explicit expression for ${}^{(3)}\Gamma_{abc}$ in terms of the object of anholonomity. In the final part, we use the existence of a locally orthonormal tetrad frame corresponding to a family of observers co-moving with the Dirac matter (which arises from the Fierz identities), to generate an expression for the generic vierbein field as a function of the Dirac bilinears.

A summary of our formulation of the self-coupled Einstein-Cartan system is given in Section \ref{Summary and conclusions}, which demonstrates in principle how this system can be reduced to a set of relations between Dirac bilinears only. A glossary of the symbols we use in this paper are provided in appendix \ref{Glossary of symbols}. Further reduction and analysis of this system is left for future publications.

\section{\label{The Einstein-Cartan-Dirac equations and conventions}The Einstein-Cartan-Dirac equations and conventions}
The Einstein-Cartan-Dirac equations in curved spacetime with torsion have the form
\begin{align}
&(\rmi\gamma^{a}e_{a}{}^{\mu}(x)\nabla_{\mu}-m)\psi=0,\label{Dirac equation in curved space} \\
&R_{\mu\nu}-\frac{1}{2}g_{\mu\nu}R+\Lambda g_{\mu\nu}=8\pi GT_{\mu\nu},\label{Einstein equation} \\
&\Upsilon_{\mu\nu}{}^{\gamma}+\delta_{\mu}^{\gamma}\Upsilon_{\nu\sigma}{}^{\sigma}-\delta_{\nu}^{\gamma}\Upsilon_{\mu\sigma}{}^{\sigma}=8\pi G\Sigma_{\mu\nu}{}^{\gamma}.\label{Cartan equation}
\end{align}
The Dirac equation (\ref{Dirac equation in curved space}) governs the dynamics of the matter sector of this system; namely, the relativistic wave-like behaviour of spin-1/2 fermionic matter. The gravitational field in the presence of Dirac matter has both curvature and torsion due to the stress-energy and spin of the Dirac matter respectively; the Einstein field and Cartan torsion equations, (\ref{Einstein equation}) and (\ref{Cartan equation}), describe these relationships.

The focus of this paper is primarily on (\ref{Dirac equation in curved space}) and its explicit invertibility for irreducible components of the connection. A brief discussion of the Einstein field equation (\ref{Einstein equation}) is given in subsection \ref{Curvature Field Equations}. The utilization of the torsion field equation (\ref{Cartan equation}) to obtain further constraints on the connection is presented in subsections \ref{Torsion Field Equations} and \ref{Constraints Arising From Torsion}. Our end result will be an in-principle integration of (\ref{Dirac equation in curved space}) and (\ref{Cartan equation}) with (\ref{Einstein equation}), with the ability to express (\ref{Einstein equation}) entirely in terms of Dirac bilinears, the state densities of matter which also act as the source of gravity. Deeper analysis of the Einstein equation using the inverted form of the Dirac equation and torsion constraints is beyond the scope of this paper, and is left to future publications.

For Dirac matter, the stress-energy and spin densities are given in terms of the spinors as \cite{Goedecke1974}, \cite{HehlHeydeKerlickNester1976}
\begin{align}
&T_{\mu\nu}=\frac{\rmi}{2}[\psibar\gamma_{\mu}(\nabla_{\nu}\psi)-(\nabla_{\nu}\psibar)\gamma_{\mu}\psi], \\
&\Sigma_{\mu\nu\gamma}=\frac{\rmi}{4}\psibar\gamma_{[\mu}\gamma_{\nu}\gamma_{\gamma]}\psi.
\end{align}
Greek and Latin indices run from 0 to 3, and correspond to coordinate and locally orthonormal frames respectively. The vierbein field $e_{a}{}^{\mu}(x)$ relates these two frames locally at each point $x$, and is quadratically related to the metric, according to
\begin{equation}
g_{\mu\nu}(x)=e^{a}{}_{\mu}(x)e^{b}{}_{\nu}(x)\eta_{ab}.
\end{equation}
For the Minkowski spacetime metric we use the particle physics sign convention, whereby the signature is negative in the spatial components:
\begin{equation}
\eta_{ab}:=\rmdiag(1,-1,-1,-1).
\end{equation}

\subsection{\label{The gravitational four-vector potentials}The gravitational four-vector potentials}
For Dirac spinor fields, the covariant derivative is of the form \cite{Crawford1993}, \cite{Yepez2011}
\begin{equation}\label{Spinor Lorentz covariant derivative}
\nabla_{\mu}\psi=\partial_{\mu}\psi+\Gamma_{\mu}\psi,
\end{equation}
where in the spinor representation, the connection coefficients are
\begin{equation}\label{Spinor connection coefficients}
\Gamma_{\mu}=\frac{1}{2}\Gamma_{\mu}{}^{ab}S_{ab}=-\frac{\rmi}{2}\Gamma_{\mu}{}^{ab}\sigma_{ab}.
\end{equation}
The object $\Gamma_{\mu}{}^{ab}$ with the leading index in the world-coordinate (holonomic) frame and the other two indices in the local (anholonomic) frame, is often referred to as the \emph{spin connection}, however we shall mostly refer to it as simply the connection. The connection transforms \emph{inhomogeneously} between the coordinate and local frames, according to \cite{HehlLemkeMielke1991}
\begin{equation}\label{Connection Inhomogeneous Frame Transformation}
\Gamma_{a}{}^{bc}=e_{a}{}^{\mu}e^{b\nu}e^{c}{}_{\lambda}\Gamma_{\mu\nu}{}^{\lambda}-e_{a}{}^{\mu}e^{b\nu}\partial_{\mu}e^{c}{}_{\nu},
\end{equation}
where $\Gamma_{a}{}^{bc}=e_{a}{}^{\mu}\Gamma_{\mu}{}^{bc}$. Note that because of the intrinsic spin of the Dirac field, $\Gamma_{\mu\nu}{}^{\lambda}$ is in general \emph{asymmetric} in $\mu,\nu$, resulting in a non-vanishing spacetime torsion \cite{Kibble1961}, \cite{BlagojevicHehl2013}. The infinitesimal Lorentz generators in the Dirac spinor representation are
\begin{equation}\label{Lorentz generators, spinor representation}
S_{ab}=-\frac{\rmi}{2}\sigma_{ab}=\frac{1}{4}[\gamma_{a},\gamma_{b}],
\end{equation}
where $\gamma_{a}$ are the \emph{Dirac matrices}, and $\sigma_{ab}\equiv\rmi/2[\gamma_{a},\gamma_{b}]$.
Taking account of the Dirac matrix anticommutator
\begin{equation}
\{\gamma_{a},\gamma_{b}\}=2\eta_{ab},
\end{equation}
it can be shown that the right-hand side of (\ref{Lorentz generators, spinor representation}) satisfies the Lie bracket identity for Lorentz generators
\begin{equation}\label{Lorentz generators, Lie bracket}
[S_{ab},S_{cd}]=\eta_{ad}S_{bc}+\eta_{bc}S_{ad}-\eta_{ac}S_{bd}-\eta_{bd}S_{ac}.
\end{equation}
Using (\ref{Spinor connection coefficients}) and (\ref{Lorentz generators, spinor representation}), we can rewrite the covariant derivative of the Dirac spinor as
\begin{equation}
\nabla_{\mu}\psi=\partial_{\mu}\psi+\frac{1}{8}\Gamma_{\mu}{}^{ab}[\gamma_{a},\gamma_{b}]\psi.
\end{equation}
Substituting this into (\ref{Dirac equation in curved space}), then absorbing the vierbeins and rearranging, the Dirac equation becomes
\begin{equation}\label{Dirac equation with spin connection}
\frac{\rmi}{8}\Gamma^{abc}\gamma_{a}[\gamma_{b},\gamma_{c}]\psi=-(\rmi\gamma^{a}\partial_{a}-m)\psi,
\end{equation}
with $\gamma^{a}\partial_{a}\equiv\gamma^{a}e_{a}{}^{\mu}\partial_{\mu}$. Using the Dirac algebra identity
\begin{equation}
\gamma_{a}\gamma_{b}\gamma_{c}=\eta_{ab}\gamma_{c}+\eta_{bc}\gamma_{a}-\eta_{ac}\gamma_{b}-\rmi\epsilon_{abcd}\gamma_{5}\gamma^{d},
\end{equation}
we can write the commutator in the last two indices as
\begin{equation}
\gamma_{a}[\gamma_{b},\gamma_{c}]=2(\eta_{ab}\gamma_{c}-\eta_{ac}\gamma_{b}-\rmi\epsilon_{abcd}\gamma_{5}\gamma^{d}).
\end{equation}
The conventions we use for $\gamma_{5}$ and the Levi-Civita symbol are those of Itzykson and Zuber \cite{ItzyksonZuber1980}:
\begin{align}\label{cases}
\epsilon^{abcd}=-\epsilon_{abcd}=\begin{cases}+1&$if $\{a,b,c,d\}$ even$ \\
-1&$odd$ \\
0&$otherwise,$\end{cases} \\
\gamma^{5}=\gamma_{5}=-(\rmi/4!)\epsilon_{\mu\nu\rho\sigma}\gamma^{\mu}\gamma^{\nu}\gamma^{\rho}\gamma^{\sigma}=\rmi\gamma^{0}\gamma^{1}\gamma^{2}\gamma^{3}=-\rmi\gamma_{0}\gamma_{1}\gamma_{2}\gamma_{3}.
\end{align}
The left-hand side operator of (\ref{Dirac equation with spin connection}) therefore becomes
\begin{align}
\frac{\rmi}{8}\Gamma^{abc}\gamma_{a}[\gamma_{b},\gamma_{c}]=\frac{\rmi}{4}\Gamma^{abc}(\eta_{ab}\gamma_{c}-\eta_{ac}\gamma_{b}-\rmi\epsilon_{abcd}\gamma_{5}\gamma^{d}) \nonumber \\
\qquad=\frac{\rmi}{4}(\eta_{ab}\eta_{cd}-\eta_{ac}\eta_{bd})\Gamma^{abc}\gamma^{d}+\frac{1}{4}\epsilon_{abcd}\Gamma^{abc}\gamma_{5}\gamma^{d} \nonumber \\
\qquad=\Omega_{d}\gamma^{d}+\Omega_{5d}\gamma_{5}\gamma^{d},
\end{align}
where we define the \emph{gravitational vector potentials} as
\begin{align}
\Omega_{d}:=\frac{1}{4}\delta_{adbc}\Gamma^{abc}=\frac{\rmi}{2}\Gamma_{c}{}^{c}{}_{d},\label{Omega definition} \\
\Omega_{5d}:=\frac{1}{4}\epsilon_{abcd}\Gamma^{abc},\label{Omega5 definition}
\end{align}
with the mixed symmetry imaginary Sylvester tensor \cite{Zund1976}
\begin{equation}\label{Delta definition}
\delta_{abcd}:=\rmi(\eta_{ac}\eta_{bd}-\eta_{ad}\eta_{bc}),
\end{equation}
playing a dual role to the Levi-Civita tensor.

\subsection{\label{Connection Irreducible Decomposition}Connection Irreducible Decomposition}
From the definitions (\ref{Omega definition}) and (\ref{Omega5 definition}), we can see that the gravitational vector potentials correspond to components of the connection $\Gamma_{abc}$. Now, since the connection corresponds to a rank-3 representation of the local Lorentz group $SO(1,3)$, we can write it as the sum of three irreducible components:
\begin{equation}
\Gamma_{abc}={}^{(1)}\Gamma_{abc}+{}^{(2)}\Gamma_{abc}+{}^{(3)}\Gamma_{abc}.
\end{equation}
Due to the antisymmetry of the connection in its second and third indices in the local frame, this irreducible decomposition can equivalently be written in terms of Young patterns as
\begin{equation}
[1]\otimes[11]=[1]\oplus[111]\oplus[21],
\end{equation}
corresponding to $(1)$ a trace term, $(2)$ a fully antisymmetric term, and $(3)$ a traceless mixed-symmetry term respectively. Written in terms of the connection, the irreducible parts are
\begin{subequations}
\begin{align}
{}^{(1)}\Gamma_{abc}&=\frac{1}{3}\eta_{ab}\Gamma_{d}{}^{d}{}_{c}-\frac{1}{3}\eta_{ac}\Gamma_{d}{}^{d}{}_{b}=-\frac{\rmi}{3}\delta_{aebc}\Gamma_{d}{}^{de}, \\
{}^{(2)}\Gamma_{abc}&=\frac{1}{3}(\Gamma_{abc}+\Gamma_{bca}+\Gamma_{cab}), \\
{}^{(3)}\Gamma_{abc}&=\frac{1}{3}(2\Gamma_{abc}-\Gamma_{bca}-\Gamma_{cab})+\frac{\rmi}{3}\delta_{aebc}\Gamma_{d}{}^{de}.
\end{align}
\end{subequations}Using (\ref{Omega definition}) and (\ref{Omega5 definition}), we can express two of the three irreducible components of the connection in terms of the gravitational four-vector ``potentials'':
\begin{subequations}
\begin{align}
{}^{(1)}\Gamma_{abc}&=-\frac{2}{3}\delta_{adbc}\Omega^{d}, \\
{}^{(2)}\Gamma_{abc}&=-\frac{2}{3}\epsilon_{abcd}\Omega_{5}^{d},
\end{align}
\end{subequations}The connection can now be written as
\begin{equation}\label{Spin connection in terms of Omegas}
\Gamma_{abc}=-\frac{2}{3}\delta_{adbc}\Omega^{d}-\frac{2}{3}\epsilon_{abcd}\Omega_{5}^{d}+{}^{(3)}\Gamma_{abc}.
\end{equation}
We have thus obtained an expression for the connection in the local frame, which allows for its replacement in terms of the bilinear Dirac matter states via the inverted forms of the Dirac equation (\ref{Inverted for Omega with closed bilinears}) and (\ref{Inverted for Omega5 with closed bilinears}), with the exception of the residual term ${}^{(3)}\Gamma_{abc}$. As we shall see in subsection \ref{Constraints Arising From Torsion}, ${}^{(3)}\Gamma_{abc}$ can be replaced by the irreducible traceless mixed-symmetry component of the object of anholonomity (\ref{Object of anholonomity}), which itself can be replaced by Dirac bilinears when the vierbein is chosen to be the bilinear tetrad (\ref{Bilinear tetrad}). Thus, we will be able to obtain an expression for the connection entirely in terms of Dirac bilinears.

\subsection{Charge conjugation and comparison with electromagnetism}
In terms of the $\Omega$-potentials, the Dirac equation now reads
\begin{equation}\label{Curved spacetime Dirac equation}
\Omega_{a}\gamma^{a}\psi+\Omega_{5a}\gamma_{5}\gamma^{a}\psi=-(\rmi\gamma^{a}\partial_{a}-m)\psi.
\end{equation}
According to the gauging of the Poincar\'{e} group \cite{BlagojevicHehl2013}, there are four translation-type potentials $\theta^{a}=e_{\mu}{}^{a}dx^{\mu}$ and six Lorentz-type potentials $\Gamma^{ab}=\Gamma_{\mu}{}^{ab}dx^{\mu}$. By introducing the two new four-vector potentials $\Omega^{a}$ and $\Omega_{5}^{a}$ to replace irreducible parts of the connection, we have increased the number of components from $6$ to $4+4=8$. The inverted Dirac equation provides two explicit expressions for these potentials, (\ref{Inverted for Omega with closed bilinears}) and (\ref{Inverted for Omega5 with closed bilinears}), that reduce the number of independent connection components back down to six.

Equation (\ref{Curved spacetime Dirac equation}) can be compared with the electromagnetically covariant Dirac equation in \emph{flat} spacetime
\begin{equation}
-qA_{a}\gamma^{a}\psi=-(\rmi\gamma^{a}\partial_{a}-m)\psi.
\end{equation}
We can see that there is an analogy between $\Omega_{a}$ and $-qA_{a}$, in the sense that these terms are coupled to $\gamma^{a}\psi$. However, in electromagnetism there is no equivalent potential to $\Omega_{5a}$, say $-qA_{5a}$, which couples to $\gamma_{5}\gamma^{a}$. Such an analogous term could in principle arise in an Abelian chiral generalization of the electromagnetic gauge group, such as local $U(1)_{L}\times U(1)_{R}$ symmetry. It is of interest to note that if the torsion field equation is taken into account in (\ref{Curved spacetime Dirac equation}), say by directly substituting the constraint (\ref{Torsion equation Omega_5 constraint}), the left-hand side of (\ref{Curved spacetime Dirac equation}) becomes non-linear in the spinors (via the axial vector term $k^{a}$), and the Hehl-Datta equation is obtained \cite{HehlDatta1971}. However, as our emphasis is on the explicit \emph{inversion} of the Dirac equation for the $\Omega$ and $\Omega_{5}$ ``potentials'', we shall treat these objects equally at this stage.

In order to proceed with the inversion process, we require the Dirac equation for the charge conjugated spinor. It can be shown \cite{Pollock2010} that in the absence of electromagnetic fields, this equation has exactly the same form as (\ref{Dirac equation in curved space}) and (\ref{Curved spacetime Dirac equation}), such that
\begin{equation}
(\rmi\gamma^{a}e_{a}{}^{\mu}(x)\nabla_{\mu}-m)\psi^{\rmc}=0,
\end{equation}
and therefore
\begin{equation}\label{Charge conjugate Dirac equation in curved spacetime, for left-mult}
\Omega_{a}\gamma^{a}\psi^{\rmc}+\Omega_{5a}\gamma_{5}\gamma^{a}\psi^{\rmc}=-(\rmi\gamma^{a}\partial_{a}-m)\psi^{\rmc}.
\end{equation}
Incidentally, in the electromagnetic case, the sign of the term carrying the charge coupling constant $q$ changes sign under a charge conjugation:
\begin{equation}
+qA_{a}\gamma^{a}\psi^{\rmc}=-(\rmi\gamma^{a}\partial_{a}-m)\psi^{\rmc}.
\end{equation}

\section{\label{The inversion procedure}The inversion procedure}
The inversion of the Dirac equation for the components of the connection which couple to spin-1/2 fermions proceeds in a similar fashion to the analogous $U(1)$ electromagnetic \cite{InglisJarvis2014} and non-Abelian $SU(2)$ \cite{InglisJarvis2012} cases. In all of these cases, the procedure involves the formation of spinor bilinears, which in the tradition of Takabayasi \cite{Takabayasi1957}, Zhelnorovich \cite{Zhelnorovich1965}, and Halbwachs \cite{Halbwachs1960}, we can write as a set of 16 tensor fields: scalar, pseudoscalar, four-vector, axial four-vector, and rank-2 tensor
\begin{subequations}
\begin{align}
\sigma=\psibar\psi,\label{Bilinear scalar definition} \\
\omega=\psibar\gamma_{5}\psi, \\
j^{a}=\psibar\gamma^{a}\psi, \\
k^{a}=\psibar\gamma_{5}\gamma^{a}\psi, \\
s^{ab}=\psibar\sigma^{ab}\psi.\label{Bilinear rank-2 tensor definition}
\end{align}
\end{subequations}
In addition, we also have the dual rank-2 tensor
\begin{equation}
\sdual^{ab}=\frac{\rmi}{2}\epsilon^{abcd}s_{cd}=\psibar\gamma_{5}\sigma^{ab}\psi,
\end{equation}
as well as two four-vectors comprising real and imaginary parts of a complex bilinear
\begin{subequations}
\begin{align}
&m^{a}+\rmi n^{a}=\psibar{}^{\rmc}\gamma^{a}\psi \\
&m^{a}=\rmRe\{\psibar{}^{\rmc}\gamma^{a}\psi\}=\frac{1}{2}(\psibar{}^{\rmc}\gamma^{a}\psi+\psibar\gamma^{a}\psi{}^{\rmc}) \\
&n^{a}=\rmIm\{\psibar{}^{\rmc}\gamma^{a}\psi\}=\frac{\rmi}{2}(\psibar\gamma^{a}\psi{}^{\rmc}-\psibar{}^{\rmc}\gamma^{a}\psi),\label{Definition of bilinear n^a}
\end{align}
\end{subequations}where $\psibar$ and $\psi^{\rmc}$ are the Dirac and charge conjugate spinors respectively. The bilinear set are all \emph{real}, except for $\omega$ and $\sdual^{ab}$, which are pure imaginary; this is just a choice of convention, which can be altered by defining the new \emph{real} bilinears $-\rmi\omega$ and $-\rmi\sdual^{ab}$. Now, left-multiplying (\ref{Curved spacetime Dirac equation}) by $\psibar\gamma^{b}$ gives
\begin{equation}
\Omega_{a}\psibar\gamma^{b}\gamma^{a}\psi+\Omega_{5a}\psibar\gamma^{b}\gamma_{5}\gamma^{a}\psi=-\rmi\psibar\gamma^{b}\gamma^{a}(\partial_{a}\psi)+m\psibar\gamma^{b}\psi.
\end{equation}
Applying the Dirac algebra identities
\begin{subequations}
\begin{align}
&\gamma^{b}\gamma^{a}=\eta^{ba}-\rmi\sigma^{ba},\label{Dirac identity gamma_b gamma_a} \\
&\{\gamma_{5},\gamma^{a}\}=0,\label{Dirac identity anticommutator gamma_5 gamma_a}
\end{align}
\end{subequations}and writing closed-form bilinears in tensor notation, we get
\begin{equation}
(\sigma\eta^{ba}-\rmi s^{ba})\Omega_{a}+(-\omega\eta^{ba}+\rmi\sdual^{ba})\Omega_{5a}=-\rmi\psibar(\partial^{b}\psi)-\psibar\sigma^{ba}(\partial_{a}\psi)+mj^{b}.\label{Bilinearized first set, first case}
\end{equation}
Likewise, left-multiplying (\ref{Curved spacetime Dirac equation}) by $\psibar\gamma_{5}\gamma^{b}$, and applying the same Dirac algebra identities yields
\begin{equation}
(\omega\eta^{ba}-\rmi\sdual^{ba})\Omega_{a}+(-\sigma\eta^{ba}+\rmi s^{ba})\Omega_{5a}=-\rmi\psibar\gamma_{5}(\partial^{b}\psi)-\psibar\gamma_{5}\sigma^{ba}(\partial_{a}\psi)+mk^{b}.\label{Bilinearized first set, second case}
\end{equation}
Following the same steps with the charge conjugate Dirac equation, left-multiplying \ref{Charge conjugate Dirac equation in curved spacetime, for left-mult} by $\psibar{}^{\rmc}\gamma^{b}$ and $\psibar{}^{\rmc}\gamma_{5}\gamma^{b}$ yields the respective equations
\begin{align}
&(\psibar{}^{\rmc}\psi{}^{\rmc}\eta^{ba}\!-\rmi\psibar{}^{\rmc}\sigma^{ba}\psi^{\rmc})\Omega_{a}\!+(-\psibar{}^{\rmc}\gamma_{5}\psi^{\rmc}\eta^{ba}\!+\rmi\psibar{}^{\rmc}\gamma_{5}\sigma^{ba}\psi{}^{\rmc})\Omega_{5a} \nonumber \\
&\qquad=-\rmi\psibar{}^{\rmc}(\partial^{b}\psi^{\rmc})-\psibar{}^{\rmc}\sigma^{ba}(\partial_{a}\psi^{\rmc})+m\psibar{}^{\rmc}\gamma^{b}\psi^{\rmc},\label{Charge conjugate bilinear Dirac equation 1} \\
&(\psibar{}^{\rmc}\gamma_{5}\psi{}^{\rmc}\eta^{ba}\!-\rmi\psibar{}^{\rmc}\gamma_{5}\sigma^{ba}\psi^{\rmc})\Omega_{a}+(-\psibar{}^{\rmc}\psi^{\rmc}\eta^{ba}\!+\rmi\psibar{}^{\rmc}\sigma^{ba}\psi{}^{\rmc})\Omega_{5a} \nonumber \\
&\qquad=-\rmi\psibar{}^{\rmc}\gamma_{5}(\partial^{b}\psi^{\rmc})-\psibar{}^{\rmc}\gamma_{5}\sigma^{ba}(\partial_{a}\psi^{\rmc})+m\psibar{}^{\rmc}\gamma_{5}\gamma^{b}\psi^{\rmc}.\label{Charge conjugate bilinear Dirac equation 2}
\end{align}
Using the definition for the charge conjugate spinor
\begin{equation}
\psi^{\rmc}=C\psibar{}^{\rmT}=\rmi\gamma^{2}\gamma^{0}\psibar{}^{\rmT},
\end{equation}
we can derive a relationship between bilinears with \emph{non-Grassmann} charge conjugate spinors and regular spinors\footnote{In the
present work we assume the spinor quantities are $c$-numbers.}
\begin{equation}\label{Charge conjugate bilinear}
\psibar{}^{\rmc}\Gamma\chi^{\rmc}=-\chibar C^{-1}\Gamma^{\rmT}C\psi,
\end{equation}
where the spinor $\chi$ may have tensor indices (ie. $\chi=\partial_{a}\psi$), and $\Gamma$ is an element of the same Dirac-Clifford algebra defining the set (\ref{Bilinear scalar definition})-(\ref{Bilinear rank-2 tensor definition}). Applying the Dirac matrix charge conjugation identities \cite{ItzyksonZuber1980}
\begin{subequations}
\begin{align}
&C^{-1}\gamma^{a\rmT}C=-\gamma^{a}, \\
&C^{-1}\gamma_{5}^{\rmT}C=\gamma_{5}, \\
&C^{-1}(\gamma_{5}\gamma^{a})^{\rmT}C=\gamma_{5}\gamma^{a}, \\
&C^{-1}\sigma^{ab\rmT}C=-\sigma^{ab}, \\
&C^{-1}(\gamma_{5}\sigma^{ab})^{\rmT}C=-\gamma_{5}\sigma^{ab},
\end{align}
\end{subequations}
we can rewrite (\ref{Charge conjugate bilinear Dirac equation 1}) and (\ref{Charge conjugate bilinear Dirac equation 2}) as
\begin{align}
&(-\sigma\eta^{ba}-\rmi s^{ba})\Omega_{a}+(\omega\eta^{ba}+\rmi\sdual^{ba})\Omega_{5a}=\rmi(\partial^{b}\psibar)\psi-(\partial_{a}\psibar)\sigma^{ba}\psi+mj^{b},\label{Bilinearized first set, third case} \\
&(-\omega\eta^{ba}-\rmi\sdual^{ba})\Omega_{a}+(\sigma\eta^{ba}+\rmi s^{ba})\Omega_{5a}=\rmi(\partial^{b}\psibar)\gamma_{5}\psi-(\partial_{a}\psibar)\gamma_{5}\sigma^{ba}\psi-mk^{b}\label{Bilinearized first set, fourth case}
\end{align}
respectively. Subtracting (\ref{Bilinearized first set, third case}) from (\ref{Bilinearized first set, first case}), and (\ref{Bilinearized first set, fourth case}) from (\ref{Bilinearized first set, third case}), yields the respective equations
\begin{align}
&2\sigma\Omega^{a}-2\omega\Omega_{5}^{a}=-\rmi\partial^{a}\sigma-[\psibar\sigma^{ab}(\partial_{b}\psi)-(\partial_{b}\psibar)\sigma^{ab}\psi],\label{Bilinearized second set, first case} \\
&2\omega\Omega^{a}-2\sigma\Omega_{5}^{a}=-\rmi\partial^{a}\omega-[\psibar\gamma_{5}\sigma^{ab}(\partial_{b}\psi)-(\partial_{b}\psibar)\gamma_{5}\sigma^{ab}\psi]+2mk^{a},\label{Bilinearized second set, second case}
\end{align}
where we have relabelled the indices. Multiplying (\ref{Bilinearized second set, first case}) and (\ref{Bilinearized second set, second case}) by $(\sigma,\omega)$ and $(\omega,\sigma)$ respectively, then subtracting the second equation from the first gives
\begin{align}
\Omega^{a}={}&\frac{1}{2}(\sigma^{2}-\omega^{2})^{-1}\{\rmi[\omega(\partial^{a}\omega)-\sigma(\partial^{a}\sigma)]+\omega[\psibar\gamma_{5}\sigma^{ab}(\partial_{b}\psi)-(\partial_{b}\psibar)\gamma_{5}\sigma^{ab}\psi] \nonumber \\
&\ -\sigma[\psibar\sigma^{ab}(\partial_{b}\psi)-(\partial_{b}\psibar)\sigma^{ab}\psi]-2m\omega k^{a}\},\label{Inverted for Omega} \\
\Omega_{5}^{a}={}&\frac{1}{2}(\sigma^{2}-\omega^{2})^{-1}\{\rmi[\sigma(\partial^{a}\omega)-\omega(\partial^{a}\sigma)]+\sigma[\psibar\gamma_{5}\sigma^{ab}(\partial_{b}\psi)-(\partial_{b}\psibar)\gamma_{5}\sigma^{ab}\psi] \nonumber \\
&\ -\omega[\psibar\sigma^{ab}(\partial_{b}\psi)-(\partial_{b}\psibar)\sigma^{ab}\psi]-2m\sigma k^{a}\},\label{Inverted for Omega5}
\end{align}
the inverted form of the Dirac equation in curved spacetime.

\section{\label{Bilinear refinement using Fierz identities}Bilinear refinement using Fierz identities}
It is apparent however, that the bracketed second and third terms in (\ref{Inverted for Omega}) and (\ref{Inverted for Omega5}) are not closed-form bilinears, due to the minus sign preventing a simple application of the Leibniz rule for derivatives. It is possible to show through a very lengthy algebraic process that Fierz expansions can be used to re-write these terms in closed tensor form. Due to the sheer length and tediousness of these calculations, they are not given here, however their derivation follows a similar process to Appendix C in \cite{InglisJarvis2014} and Appendix B in \cite{InglisJarvis2016}.

The Fierz expansion can be used to write the outer product of two spinors $\psi\chibar$, which is a $4\times 4$ matrix in the spinor degrees of freedom, as a sum of terms over the basis of Dirac-Clifford matrices with bilinear coefficients
\begin{equation}\label{Fierz expansion}
\psi\chibar=\frac{1}{4}(\chibar\psi)I+\frac{1}{4}(\chibar\gamma_{a}\psi)\gamma^{a}+\frac{1}{8}(\chibar\sigma_{ab}\psi)\sigma^{ab}-\frac{1}{4}(\chibar\gamma_{5}\gamma_{a}\psi)\gamma_{5}\gamma^{a}+\frac{1}{4}(\chibar\gamma_{5}\psi)\gamma_{5},
\end{equation}
which can be derived from the more formal expression
\begin{equation}\label{Formal Fierz expansion}
\psi\chibar=\sum_{R=1}^{16}a_{R}\Gamma_{R}
\end{equation}
where $R=1,...,16$ runs over all of the elements of the Dirac-Clifford algebra. Multiplying (\ref{Formal Fierz expansion}) from the right by Dirac matrix $\Gamma_{B}$ [where $B$ runs over the types: \emph{scalar}, ..., \emph{rank-2 tensor} in (\ref{Bilinear scalar definition})-(\ref{Bilinear rank-2 tensor definition})], and using the trace identities
\begin{align}
\rmTr(\Gamma_{R}\Gamma_{B})=\begin{cases}\rmTr(\Gamma_{B}^{2})&$if $R=B,\\
0&$otherwise,$\end{cases}
\end{align}
\begin{equation}
\rmTr(\psi\chibar\Gamma_{B})=\chibar\Gamma_{B}\psi,\label{Bilinear trace identity}
\end{equation}
along with the trace properties of the Dirac matrices, one can easily derive (\ref{Fierz expansion}).

Following a very tedious process of applying (\ref{Fierz expansion}) to the terms in (\ref{Inverted for Omega}) and (\ref{Inverted for Omega5}) where the spinors are visible, we obtain the purely bilinear expressions
\begin{align}
&\omega[\psibar\gamma_{5}\sigma^{ab}(\partial_{b}\psi)-(\partial_{b}\psibar)\gamma_{5}\sigma^{ab}\psi]-\sigma[\psibar\sigma^{ab}(\partial_{b}\psi)-(\partial_{b}\psibar)\sigma^{ab}\psi] \nonumber \\
&\ =(\sigma^{2}-\omega^{2})^{-1}\{s^{ab}[\omega j^{c}(\partial_{b}k_{c})+\rmi\sigma m^{c}(\partial_{b}n_{c})]-\sdual^{ab}[\sigma j^{c}(\partial_{b}k_{c})+\rmi\omega m^{c}(\partial_{b}n_{c})]\} \nonumber \\
&\qquad\ +\delta^{abcd}[k_{c}(\partial_{b}k_{d})-j_{c}(\partial_{b}j_{d})],\label{Main Fierz identitity with rank-2 tensors 1} \\
&\sigma[\psibar\gamma_{5}\sigma^{ab}(\partial_{b}\psi)-(\partial_{b}\psibar)\gamma_{5}\sigma^{ab}\psi]-\omega[\psibar\sigma^{ab}(\partial_{b}\psi)-(\partial_{b}\psibar)\sigma^{ab}\psi] \nonumber \\
&\ =(\sigma^{2}-\omega^{2})^{-1}\{s^{ab}[\sigma j^{c}(\partial_{b}k_{c})+\rmi\omega m^{c}(\partial_{b}n_{c})]-\sdual^{ab}[\omega j^{c}(\partial_{b}k_{c})+\rmi\sigma m^{c}(\partial_{b}n_{c})]\} \nonumber \\
&\qquad\ +\epsilon^{abcd}[k_{c}(\partial_{b}k_{d})-j_{c}(\partial_{b}j_{d})].\label{Main Fierz identitity with rank-2 tensors 2}
\end{align}
Using the Fierz identities \cite{Crawford1985}
\begin{subequations}
\begin{align}
&s^{ab}=(\sigma^{2}-\omega^{2})^{-1}(\sigma\epsilon^{abcd}-\omega\delta^{abcd})j_{c}k_{d} \\
&\sdual^{ab}=(\sigma^{2}-\omega^{2})^{-1}(\omega\epsilon^{abcd}-\sigma\delta^{abcd})j_{c}k_{d}, \\
&\rmi\epsilon^{abcd}j_{c}k_{d}=\rmi(m^{a}n^{b}-m^{b}n^{a})=\delta^{abcd}m_{c}n_{d}, \\
&\rmi\delta^{abcd}j_{c}k_{d}=-j^{a}k^{b}+j^{b}k^{a}=\epsilon^{abcd}m_{c}n_{d},
\end{align}
\end{subequations}
the expressions within the curved braces can be recast as
\begin{align}
&s^{ab}[\omega j^{c}(\partial_{b}k_{c})+\rmi\sigma m^{c}(\partial_{b}n_{c})]-\sdual^{ab}[\sigma j^{c}(\partial_{b}k_{c})+\rmi\omega m^{c}(\partial_{b}n_{c})] \nonumber \\
&\qquad=\delta^{abcd}[j_{c}j^{e}k_{d}(\partial_{b}k_{e})+m_{c}m^{e}n_{d}(\partial_{b}n_{e})],\label{Curved bracket contents 1} \\
&s^{ab}[\sigma j^{c}(\partial_{b}k_{c})+\rmi\omega m^{c}(\partial_{b}n_{c})]-\sdual^{ab}[\omega j^{c}(\partial_{b}k_{c})+\rmi\sigma m^{c}(\partial_{b}n_{c})] \nonumber \\
&\qquad=\epsilon^{abcd}[j_{c}j^{e}k_{d}(\partial_{b}k_{e})+m_{c}m^{e}n_{d}(\partial_{b}n_{e})].\label{Curved bracket contents 2}
\end{align}
To proceed further, we require the tetrad frame of four-vector bilinears, with scalar normalizing factor:
\begin{equation}\label{Bilinear tetrad}
t_{\alpha}{}^{a}=(\sigma^{2}-\omega^{2})^{-1/2}[j^{a},m^{a},n^{a},k^{a}],
\end{equation}
where $\alpha=0,1,2,3$ labels the columns. The details of this local frame are discussed in subsection \ref{The Dirac Bilinear Local Frame}. The tetrad orthonormality implies
\begin{equation}\label{Metric tetrad identity}
t_{\alpha}{}^{a}t^{\alpha}{}_{b}=\delta^{a}{}_{b}=(\sigma^{2}-\omega^{2})^{-1}(j^{a}j_{b}-m^{a}m_{b}-n^{a}n_{b}-k^{a}k_{b}),
\end{equation}
and taking the derivative yields
\begin{equation}\label{Negative derivative tetrad identity}
t_{\alpha}{}^{a}(\partial_{b}t_{\beta a})=-t_{\beta}{}^{a}(\partial_{b}t_{\alpha a}),
\end{equation}
which provides the freedom to switch what bilinear the derivative operator acts on, when the Lorentz index is summed over. In the special case where $\alpha=\beta$, we can replace four-vectors entirely via
\begin{equation}\label{Invariant length squared derivative}
j^{a}(\partial_{b}j_{a})=-m^{a}(\partial_{b}m_{a})=-n^{a}(\partial_{b}n_{a})=-k^{a}(\partial_{b}k_{a})=\sigma(\partial_{b}\sigma)-\omega(\partial_{b}\omega),
\end{equation}
which is just the derivative of the invariant length squared Fierz identity \cite{Crawford1985}. Note that (\ref{Invariant length squared derivative}) is consistent with (\ref{Metric tetrad identity}), when setting $a=b$ and summing. Applying these identities to the square brackets in (\ref{Curved bracket contents 1}) and (\ref{Curved bracket contents 2}) gives, after some manipulation
\begin{align}
&j_{c}j^{e}k_{d}(\partial_{b}k_{e})+m_{c}m^{e}n_{d}(\partial_{b}n_{e}) \nonumber \\
&\qquad=\frac{1}{2}(\sigma^{2}-\omega^{2})[j_{c}(\partial_{b}j_{d})-k_{c}(\partial_{b}k_{d})+n_{c}(\partial_{b}n_{d})+m_{c}(\partial_{b}m_{d})].
\end{align}
We now write a much simpler form of (\ref{Main Fierz identitity with rank-2 tensors 1}) and (\ref{Main Fierz identitity with rank-2 tensors 2}):
\begin{align}
&\omega[\psibar\gamma_{5}\sigma^{ab}(\partial_{b}\psi)-(\partial_{b}\psibar)\gamma_{5}\sigma^{ab}\psi]-\sigma[\psibar\sigma^{ab}(\partial_{b}\psi)-(\partial_{b}\psibar)\sigma^{ab}\psi] \nonumber \\
&\qquad=\frac{1}{2}\delta^{abcd}[-j_{c}(\partial_{b}j_{d})+k_{c}(\partial_{b}k_{d})+n_{c}(\partial_{b}n_{d})+m_{c}(\partial_{b}m_{d})],\label{Main Fierz identitity 1} \\
&\sigma[\psibar\gamma_{5}\sigma^{ab}(\partial_{b}\psi)-(\partial_{b}\psibar)\gamma_{5}\sigma^{ab}\psi]-\omega[\psibar\sigma^{ab}(\partial_{b}\psi)-(\partial_{b}\psibar)\sigma^{ab}\psi] \nonumber \\
&\qquad=\frac{1}{2}\epsilon^{abcd}[-j_{c}(\partial_{b}j_{d})+k_{c}(\partial_{b}k_{d})+n_{c}(\partial_{b}n_{d})+m_{c}(\partial_{b}m_{d})].\label{Main Fierz identitity 2}
\end{align}
Finally, substituting into (\ref{Inverted for Omega}) and (\ref{Inverted for Omega5}), we obtain the gravitational four-vector potentials in terms of \emph{closed-form bilinears} only
\begin{align}
\Omega^{a}={}&\frac{1}{2}(\sigma^{2}-\omega^{2})^{-1}\{\rmi[\omega(\partial^{a}\omega)-\sigma(\partial^{a}\sigma)]-2m\omega k^{a} \nonumber \\
&+\frac{1}{2}\delta^{abcd}[-j_{c}(\partial_{b}j_{d})+k_{c}(\partial_{b}k_{d})+n_{c}(\partial_{b}n_{d})+m_{c}(\partial_{b}m_{d})]\},\label{Inverted for Omega with closed bilinears} \\
\Omega_{5}^{a}={}&\frac{1}{2}(\sigma^{2}-\omega^{2})^{-1}\{\rmi[\sigma(\partial^{a}\omega)-\omega(\partial^{a}\sigma)]-2m\sigma k^{a} \nonumber \\
&+\frac{1}{2}\epsilon^{abcd}[-j_{c}(\partial_{b}j_{d})+k_{c}(\partial_{b}k_{d})+n_{c}(\partial_{b}n_{d})+m_{c}(\partial_{b}m_{d})]\}.\label{Inverted for Omega5 with closed bilinears}
\end{align}
Comparing with the inverted Dirac equation in the electromagnetic case \cite{InglisJarvis2014}
\begin{align}
A^{a}={}&\frac{1}{2q}(\sigma^{2}-\omega^{2})^{-1}\{\epsilon^{abcd}[j_{c}(\partial_{b}k_{d})-k_{c}(\partial_{b}j_{d})]+m^{b}(\partial^{a}n_{b})-2m\sigma j^{a}\} \nonumber \\
&+\frac{1}{2q}(\sigma^{2}-\omega^{2})^{-2}\{\delta^{abcd}j_{c}k_{d}[\omega(\partial_{b}\sigma)-\sigma(\partial_{b}\omega)] \nonumber \\
&\qquad\qquad\qquad\qquad\qquad\qquad\qquad+\epsilon^{abcd}j_{c}k_{d}[\omega(\partial_{b}\omega)-\sigma(\partial_{b}\sigma)]\},
\end{align}
where the totality of the $U(1)$ gauge dependence is represented by the $m^{b}(\partial^{a}n_{b})$ term, we can see some apparent structural similarities, despite their differences.

\section{\label{The Einstein-Cartan-Dirac self-coupled system}The Einstein-Cartan-Dirac self-coupled system}

\subsection{\label{Curvature Field Equations}Curvature Field Equations}
Consider Einstein's equations coupled to a source term with generally non-vanishing cosmological constant:
\begin{equation}\label{Einstein's field equations}
R_{\mu\nu}-\frac{1}{2}g_{\mu\nu}R+\Lambda g_{\mu\nu}=8\pi GT_{\mu\nu}.
\end{equation}
In the present case, where the gravitational field couples to the Dirac field, the asymmetric canonical stress-energy tensor on the right hand side is given by \cite{Goedecke1974}
\begin{equation}\label{Dirac bilinear stress-energy tensor}
T_{\mu\nu}=\frac{\rmi}{2}[\psibar\gamma_{\mu}(\nabla_{\nu}\psi)-(\nabla_{\nu}\psibar)\gamma_{\mu}\psi].
\end{equation}
This can be rewritten in terms of Dirac bilinears with the use of Fierz identities \cite{InglisJarvis2016}, which yields
\begin{equation}
T_{\mu\nu}=\frac{1}{2}(\sigma^{2}-\omega^{2})^{-1}[\rmi k_{\mu}(\omega\partial_{\nu}\sigma-\sigma\partial_{\nu}\omega)-g^{-1/2}\epsilon_{\mu\sigma\rho\epsilon}(\nabla_{\nu}j^{\sigma})j^{\rho}k^{\epsilon}+j_{\mu}m^{\sigma}(\nabla_{\nu}n_{\sigma})].\label{Dirac matter stress-energy tensor}
\end{equation}
One the other side of the equation, we have the contractions of the curvature tensor, which in terms of the spin connection is given by \cite{Kibble1961}, \cite{AliCafaroCapozzielloCorda2009}
\begin{equation}
R^{a}{}_{b\mu\nu}=\partial_{\nu}\Gamma_{\mu}{}^{a}{}_{b}-\partial_{\mu}\Gamma_{\nu}{}^{a}{}_{b}-\Gamma_{\mu}{}^{a}{}_{e}\Gamma_{\nu}{}^{e}{}_{b}+\Gamma_{\nu}{}^{a}{}_{e}\Gamma_{\mu}{}^{e}{}_{b}.
\end{equation}
It is important to note that the curvature tensor we use is not the Riemannian one from standard general relativity, due to the presence of a non-vanishing torsion field. The non-Riemannian component of $R^{a}{}_{b\mu\nu}$ vanishes in the limit where the torsion vanishes. An expression in terms of locally orthonormal components is obtained, as usual, via contraction with the vierbein
\begin{align}
&R^{ab}{}_{cd}\equiv e^{\mu}{}_{c}e^{\nu}{}_{d}R^{ab}{}_{\mu\nu} \nonumber \\
&\ =[e^{\mu}{}_{c}(\partial_{d}e_{\mu}{}^{e})-e^{\mu}{}_{d}(\partial_{c}e_{\mu}{}^{e})]\Gamma_{e}{}^{ab}+\partial_{d}\Gamma_{c}{}^{ab}-\partial_{c}\Gamma_{d}{}^{ab}-\Gamma_{c}{}^{a}{}_{e}\Gamma_{d}{}^{eb}+\Gamma_{d}{}^{a}{}_{e}\Gamma_{c}{}^{eb}.
\end{align}
Switching the derivatives on the vierbein terms (which reverses the sign), we can write the curvature tensor as
\begin{equation}
R^{ab}{}_{cd}=\Theta^{e}{}_{cd}\Gamma_{e}{}^{ab}+\partial_{d}\Gamma_{c}{}^{ab}-\partial_{c}\Gamma_{d}{}^{ab}-\Gamma_{c}{}^{a}{}_{e}\Gamma_{d}{}^{eb}+\Gamma_{d}{}^{a}{}_{e}\Gamma_{c}{}^{eb},
\end{equation}
where we define the \emph{objects of anholonomity} as
\begin{equation}\label{Object of anholonomity}
\Theta_{abc}\equiv e_{\mu a}(\partial_{b}e^{\mu}{}_{c}-\partial_{c}e^{\mu}{}_{b}),
\end{equation}
which are representative of the non-commutativity of the tetrad basis \cite{HehlHeydeKerlickNester1976}. Contracting $b$ and $d$ in the curvature tensor yields the Ricci tensor
\begin{equation}\label{Ricci tensor}
R^{a}{}_{b}=\Theta^{c}{}_{bd}\Gamma_{c}{}^{ad}+\partial_{c}\Gamma_{b}{}^{ac}-\partial_{b}\Gamma_{c}{}^{ac}-\Gamma_{b}{}^{a}{}_{c}\Gamma_{d}{}^{cd}+\Gamma_{d}{}^{a}{}_{c}\Gamma_{b}{}^{cd},
\end{equation}
with the final contraction yielding the Ricci scalar
\begin{equation}\label{Ricci scalar}
R=\Theta_{abc}\Gamma^{abc}+2\partial_{a}\Gamma_{b}{}^{ba}-\Gamma_{a}{}^{a}{}_{b}\Gamma_{c}{}^{bc}+\Gamma_{abc}\Gamma^{bca}.
\end{equation}

\subsection{\label{Torsion Field Equations}Torsion Field Equations}

The torsion tensor is defined as the degree to which the affine connection fails to be symmetric:
\begin{equation}
\Upsilon_{\mu\nu}{}^{\lambda}=\Gamma_{\mu\nu}{}^{\lambda}-\Gamma_{\nu\mu}{}^{\lambda}.
\end{equation}
A particle field with intrinsic quantum spin will act as the source of a non-vanishing torsion field, in an analogous manner to stress-energy acting as the source of curvature \cite{BlagojevicHehl2013}. The torsion field equation is given by
\begin{equation}\label{Torsion field equation}
\Upsilon_{\mu\nu}{}^{\gamma}+\delta_{\mu}^{\gamma}\Upsilon_{\nu\sigma}{}^{\sigma}-\delta_{\nu}^{\gamma}\Upsilon_{\mu\sigma}{}^{\sigma}=8\pi G\Sigma_{\mu\nu}{}^{\gamma}.
\end{equation}
Together, the curvature and torsion gravitational field equations, (\ref{Einstein's field equations}) and (\ref{Torsion field equation}), comprise the Einstein-Cartan(-Sciama-Kibble) equations.

In terms of the spinor field, the canonical spin momentum tensor in a locally orthonormal frame is
\begin{equation}\label{Dirac spin tensor, spinor form}
\Sigma^{abc}=\frac{\rmi}{4}\psibar\gamma^{[a}\gamma^{b}\gamma^{c]}\psi.
\end{equation}
Given that
\begin{equation}
\gamma^{[a}\gamma^{b}\gamma^{c]}=\frac{1}{6}(\gamma^{a}\gamma^{b}\gamma^{c}-\gamma^{a}\gamma^{c}\gamma^{b}+\gamma^{b}\gamma^{c}\gamma^{a}-\gamma^{b}\gamma^{a}\gamma^{c}+\gamma^{c}\gamma^{a}\gamma^{b}-\gamma^{c}\gamma^{b}\gamma^{a}),
\end{equation}
we can apply the Dirac algebra identities
\begin{subequations}
\begin{align}
&\gamma^{a}\gamma^{b}=2\eta^{ab}-\gamma^{b}\gamma^{a} \\
&\gamma^{a}\sigma^{bc}=\rmi\eta^{ab}\gamma^{c}-\rmi\eta^{ac}\gamma^{b}+\epsilon^{abcd}\gamma_{5}\gamma_{d}
\end{align}
\end{subequations}
to obtain
\begin{equation}
\gamma^{[a}\gamma^{b}\gamma^{c]}=-\rmi\epsilon^{abcd}\gamma_{5}\gamma_{d}.
\end{equation}
Substituting into (\ref{Dirac spin tensor, spinor form}), we find
\begin{equation}\label{Dirac spin tensor, bilinear form}
\Sigma^{abc}=\frac{1}{4}\epsilon^{abcd}k_{d},
\end{equation}
the spin angular momentum tensor of the Dirac field is proportional to the rank-3 dual of the axial vector bilinear. With regards to the left-hand side of (\ref{Torsion field equation}), using the connection transformation rule (\ref{Connection Inhomogeneous Frame Transformation}), we can write the torsion in terms of the object of anholonomity and connection
\begin{equation}\label{Torsion in terms of anholonomity and spin connection}
\Upsilon_{abc}\equiv\Upsilon_{\mu\nu}{}^{\lambda}e^{\mu}{}_{a}e^{\nu}{}_{b}e_{\lambda c}=\Theta_{cba}-\Gamma_{abc}+\Gamma_{bac}.
\end{equation}
Alternatively, taking an appropriate cyclic combination of the torsion, the connection can be written as \cite{Blagojevic2003}, \cite{AliCafaroCapozzielloCorda2009}
\begin{equation}\label{Spin connection in terms of objects of non-hol}
\Gamma_{abc}=K_{abc}+\frac{1}{2}(\Theta_{abc}-\Theta_{bca}-\Theta_{cab}),
\end{equation}
where we define the \emph{contorsion} tensor to be
\begin{equation}
K_{abc}:=\frac{1}{2}(-\Upsilon_{abc}+\Upsilon_{bca}-\Upsilon_{cab}).
\end{equation}

\subsection{\label{Constraints Arising From Torsion}Constraints Arising From Torsion}

We shall now demonstrate how the torsion field equation can be used to obtain a further, very useful set of constraints on the Einstein-Cartan-Dirac system. For convenience, we shall consider the torsion field equation in a local frame
\begin{equation}\label{Torsion field equation in local frame}
\Upsilon_{abc}+\eta_{ac}\Upsilon_{b}{}^{d}{}_{d}-\eta_{bc}\Upsilon_{a}{}^{d}{}_{d}=8\pi G\Sigma_{abc}.
\end{equation}
Now, substituting the irreducible decomposition of the connection (\ref{Spin connection in terms of Omegas}) into the torsion (\ref{Torsion in terms of anholonomity and spin connection}), we obtain
\begin{equation}\label{Torsion in terms of spin connection and anholonomity}
\Upsilon_{abc}=-\Theta_{cab}+{}^{(3)}\Gamma_{cab}-\frac{2}{3}\delta_{cdab}\Omega^{d}+\frac{4}{3}\epsilon_{cdab}\Omega_{5}^{d},
\end{equation}
where we have used the cyclic identities
\begin{align}
&\delta_{adbc}-\delta_{bdac}=-\delta_{cdab}, \\
&{}^{(3)}\Gamma_{abc}-{}^{(3)}\Gamma_{bac}=-{}^{(3)}\Gamma_{cab}.
\end{align}
Taking the trace of (\ref{Torsion in terms of spin connection and anholonomity}) in the last two indices, the ${}^{(3)}\Gamma$ and $\Omega_{5}$ terms vanish, and we find
\begin{equation}\label{Trace of torsion}
\Upsilon_{a}{}^{b}{}_{b}=\Theta_{b}{}^{b}{}_{a}+2\rmi\Omega_{a},
\end{equation}
where we have used the antisymmetry of $\Theta_{abc}$ in $bc$. Substituting (\ref{Torsion in terms of spin connection and anholonomity}), (\ref{Trace of torsion}), and (\ref{Dirac spin tensor, bilinear form}) into (\ref{Torsion field equation in local frame}), then gathering terms and rearranging, we obtain an explicit expression for the remaining component of the connection
\begin{equation}\label{w-solution of torsion equation}
{}^{(3)}\Gamma_{abc}=\Theta_{abc}+\rmi\delta_{adbc}\Theta_{e}{}^{ed}-\frac{4}{3}\delta_{adbc}\Omega^{d}-\frac{4}{3}\epsilon_{abcd}\Omega_{5}^{d}+2\pi G\epsilon_{abcd}k^{d}.
\end{equation}
Taking the trace of (\ref{w-solution of torsion equation}), the left-hand side and Levi-Civita terms vanish, and we obtain the constraint on the gravitational vector potential
\begin{equation}\label{Torsion equation Omega constraint}
\Omega_{a}=\frac{\rmi}{2}\Theta_{b}{}^{b}{}_{a}.
\end{equation}
Similarly, when we fully contract both sides of (\ref{w-solution of torsion equation}) with the Levi-Civita tensor, the left-hand side and $\delta$-dependent terms vanish. Using the Levi-Civita contraction identity
\begin{equation}
\epsilon_{abcd}\epsilon^{abcf}=-6\delta_{d}{}^{f},
\end{equation}
where the factors of $\sqrt{|g|}$ cancel out, we obtain the constraint on the dual gravitational potential
\begin{equation}\label{Torsion equation Omega_5 constraint}
\Omega_{5}^{d}=-\frac{1}{8}\Theta_{abc}\epsilon^{abcd}+\frac{3\pi}{2}Gk^{d}.
\end{equation}
Substituting our constraints (\ref{Torsion equation Omega constraint}) and (\ref{Torsion equation Omega_5 constraint}) back into (\ref{w-solution of torsion equation}), we obtain the expression
\begin{equation}\label{w_abc torsion constraint in terms of anholonomity}
{}^{(3)}\Gamma_{abc}=\frac{1}{3}(2\Theta_{abc}-\Theta_{bca}-\Theta_{cab})+\frac{\rmi}{3}\delta_{adbc}\Theta_{e}{}^{ed}.
\end{equation}
This expression can be interpreted as that due to the constraints imposed by the Cartan torsion equation (\ref{Torsion field equation in local frame}), the traceless mixed symmetry irreducible component of the connection is equal to the traceless mixed symmetry irreducible component of the object of anholonomity. Substituting the constraints (\ref{Torsion equation Omega constraint}), (\ref{Torsion equation Omega_5 constraint}), and (\ref{w_abc torsion constraint in terms of anholonomity}) into (\ref{Torsion in terms of spin connection and anholonomity}), we obtain the simple form of the torsion
\begin{equation}\label{Torsion equation torsion solution}
\Upsilon_{abc}=8\pi G\Sigma_{abc},
\end{equation}
which is obviously a solution of (\ref{Torsion field equation in local frame}) due to the vanishing trace of the fully antisymmetric spin tensor. Substituting the same constraints into the connection (\ref{Spin connection in terms of Omegas}) we obtain
\begin{equation}\label{Spin connection under torsion equation solution}
\Gamma_{abc}=-4\pi G\Sigma_{abc}+\frac{1}{2}(\Theta_{abc}-\Theta_{bca}-\Theta_{cab}),
\end{equation}
which is consistent with (\ref{Spin connection in terms of objects of non-hol}), and the contorsion solution corresponding to (\ref{Torsion equation torsion solution}):
\begin{equation}
K_{abc}=-4\pi G\Sigma_{abc}.
\end{equation}

\subsection{\label{The Dirac Bilinear Local Frame}The Dirac Bilinear Local Frame}

In section \ref{Bilinear refinement using Fierz identities}, we used the fact that there is a local orthonormal tetrad frame corresponding to a family of observers comoving with the Dirac matter. Using the Fierz identities for the four four-vector quantities\footnote{Subjecting the spinors to a complex phase transformation, causes the $m$-$n$ plane to rotate by an angle corresponding to double the phase parameter, whereas the $j$ and $k$ vectors are left invariant.} derived from the Dirac algebra \cite{Crawford1985}, \cite{Kaempffer1981}, \cite{Lounesto2001}, \cite{Takahashi1983}
\begin{align}
&j_{\mu}j^{\mu}=-m_{\mu}m^{\mu}=-n_{\mu}n^{\mu}=-k_{\mu}k^{\mu}=\sigma^{2}-\omega^{2},\label{Four-vector bilinear invariant lengths} \\
&j_{\mu}m^{\mu}=j_{\mu}n^{\mu}=j_{\mu}k^{\mu}=m_{\mu}n^{\mu}=m_{\mu}k^{\mu}=n_{\mu}k^{\mu}=0,\label{Four-vector orthogonality relations}
\end{align}
where these bilinears are defined in terms of the spinors in (\ref{Bilinear scalar definition})-(\ref{Definition of bilinear n^a}), a local tetrad frame
\begin{align}\label{Dirac tetrad frame}
(\bt_{0},\bt_{1},\bt_{2},\bt_{3}){}&=(\sigma^{2}-\omega^{2})^{-1/2}(\bj,\boldm,\bn,\bk) \nonumber \\
&=(\sigma^{2}-\omega^{2})^{-1/2}(j^{\mu},m^{\mu},n^{\mu},k^{\mu})\partial_{\mu},
\end{align}
can be constructed, with the time-like direction given by $\bj$, and the three space-like directions given by $\boldm$, $\bn$, and $\bk$, with normalising factor $(\sigma^{2}-\omega^{2})^{-1/2}$ equal to the reciprocal of the invariant length of the four-vectors via (\ref{Four-vector bilinear invariant lengths}). This bilinear tetrad is one of infinitely many local orthonormal frames, related to one another by a local Lorentz transformation. It is pertinent to ask whether we are able to describe the components of an arbitrary vierbein field in the coordinate frame $e_{a}{}^{\mu}$ in terms of the bilinears which appear in $t_{a}{}^{\mu}$. Consider an arbitrary four-vector field $\bV$ in terms of the coordinate frame $(c)$, and the bilinear $(b)$ and generic $(g)$ tetrad local frames
\begin{equation}
\bV=V^{(c)\mu}\partial_{\mu}=V^{(b)i}\bt_{i}=V^{(g)a}\be_{a}.
\end{equation}
Taking all right-hand parts of the equation with respect to the coordinate frame gives the relationship between the various components
\begin{equation}\label{Coordinate frame components in terms of frame components}
V^{(c)\mu}=t_{i}{}^{\mu}V^{(b)i}=e_{a}{}^{\mu}V^{(g)a}.
\end{equation}
Contracting (\ref{Coordinate frame components in terms of frame components}) with the inverse of the bilinear tetrad gives
\begin{equation}\label{Bilinear fram components in terms of generic frame components}
V^{(b)i}=t^{i}{}_{\mu}e_{a}{}^{\mu}V^{(g)a}=t^{i}{}_{a}V^{(g)a},
\end{equation}
Substituting (\ref{Bilinear fram components in terms of generic frame components}) into (\ref{Coordinate frame components in terms of frame components}) yields an expression for the generic vierbein frame in terms of the Dirac bilinear frame
\begin{equation}\label{Generic vierbein in terms of bilinear tetrads}
e_{a}{}^{\mu}=t_{i}{}^{\mu}t^{i}{}_{a}=(\sigma^{2}-\omega^{2})^{-1}(j_{a}j^{\mu}-m_{a}m^{\mu}-n_{a}n^{\mu}-k_{a}k^{\mu}).
\end{equation}
This expression for the generic vierbein field in terms of the Dirac bilinears provides us with a tool for calculating \emph{internal} solutions of the Einstein-Cartan-Dirac equations, in regions where the Dirac field is non-vanishing; vacuum solutions must be matched on the matter boundary. The expression (\ref{Generic vierbein in terms of bilinear tetrads}) simplifies the self-coupled equations summarized below in section \ref{Summary and conclusions} by reducing the number of fields we need to solve for, to the number of independent parameters in the bilinears. On the other hand, due to the length of the right-hand side of (\ref{Generic vierbein in terms of bilinear tetrads}) the equations may appear far more untidy. However, this will be offset by the application of Fierz identities between contracted bilinears. An explicit algebraic reduction is deferred to future publications.

\section{\label{Summary and conclusions}Summary and conclusions}

For the sake of clarity, we shall collate our main results. We have Einstein's equations
\begin{equation}\label{Summary Einstein's equations}
R_{\mu\nu}-\frac{1}{2}g_{\mu\nu}R+\Lambda g_{\mu\nu}=8\pi GT_{\mu\nu},
\end{equation}
where on the right-hand side, we have the Dirac matter stress-energy tensor
\begin{equation}
T_{\mu\nu}=\frac{1}{2}(\sigma^{2}-\omega^{2})^{-1}[\rmi k_{\mu}(\omega\partial_{\nu}\sigma-\sigma\partial_{\nu}\omega)-g^{-1/2}\epsilon_{\mu\sigma\rho\epsilon}(\nabla_{\nu}j^{\sigma})j^{\rho}k^{\epsilon}+j_{\mu}m^{\sigma}(\nabla_{\nu}n_{\sigma})],
\end{equation}
and on the left, we have the Ricci tensor and scalar, which in the local frame are respectively
\begin{align}
&R^{a}{}_{b}=\Theta^{c}{}_{bd}\Gamma_{c}{}^{ad}+\partial_{c}\Gamma_{b}{}^{ac}-\partial_{b}\Gamma_{c}{}^{ac}-\Gamma_{b}{}^{a}{}_{c}\Gamma_{d}{}^{cd}+\Gamma_{d}{}^{a}{}_{c}\Gamma_{b}{}^{cd}, \\
&R=\Theta_{abc}\Gamma^{abc}+2\partial_{a}\Gamma_{b}{}^{ba}-\Gamma_{a}{}^{a}{}_{b}\Gamma_{c}{}^{bc}+\Gamma_{abc}\Gamma^{bca}.
\end{align}
Note that our curvature terms implicitly contain a non-zero torsion component. The covariant derivatives in the stress-energy tensor contain the connection with world indices, which due to its inhomogeneous transformation law, can be written in terms of the local frame as
\begin{equation}
\Gamma_{\mu\nu}{}^{\lambda}=e^{a}{}_{\mu}e_{b\nu}e_{c}{}^{\lambda}\Gamma_{a}{}^{bc}+e^{a}{}_{\mu}e_{b}{}^{\lambda}\partial_{a}e^{b}{}_{\nu}.
\end{equation}
Reducing the connection into three irreducible terms, and taking account of the torsion equation, the connection in the local frame can be written as
\begin{equation}\label{Connection in local frame, final form}
\Gamma_{abc}=-\frac{2}{3}\delta_{adbc}\Omega^{d}-\frac{2}{3}\epsilon_{abcd}\Omega_{5}^{d}+\frac{1}{3}(2\Theta_{abc}-\Theta_{bca}-\Theta_{cab})+\frac{\rmi}{3}\delta_{adbc}\Theta_{e}{}^{ed},
\end{equation}
where the first two terms can be written in terms of Dirac bilinears, via the gravitational vector potentials obtained by inverting the Dirac equation:
\begin{align}
\Omega^{a}={}&\frac{1}{2}(\sigma^{2}-\omega^{2})^{-1}\{\rmi[\omega(\partial^{a}\omega)-\sigma(\partial^{a}\sigma)]-2m\omega k^{a} \nonumber \\
&+\frac{1}{2}\delta^{abcd}[-j_{c}(\partial_{b}j_{d})+k_{c}(\partial_{b}k_{d})+n_{c}(\partial_{b}n_{d})+m_{c}(\partial_{b}m_{d})]\},\label{Summary inverted Dirac equation Omega} \\
\Omega_{5}^{a}={}&\frac{1}{2}(\sigma^{2}-\omega^{2})^{-1}\{\rmi[\sigma(\partial^{a}\omega)-\omega(\partial^{a}\sigma)]-2m\sigma k^{a} \nonumber \\
&+\frac{1}{2}\epsilon^{abcd}[-j_{c}(\partial_{b}j_{d})+k_{c}(\partial_{b}k_{d})+n_{c}(\partial_{b}n_{d})+m_{c}(\partial_{b}m_{d})]\}.\label{Summary inverted Dirac equation Omega5}
\end{align}
The object of anholonomity is given in terms of the vierbein as
\begin{equation}
\Theta_{abc}\equiv e_{\mu a}(\partial_{b}e^{\mu}{}_{c}-\partial_{c}e^{\mu}{}_{b}).
\end{equation}
Using the fact that, due to the existence of a local orthonormal frame carried by observers co-moving with the Dirac field, the vierbein field can be written as
\begin{equation}
e_{a}{}^{\mu}=(\sigma^{2}-\omega^{2})^{-1}(j_{a}j^{\mu}-m_{a}m^{\mu}-n_{a}n^{\mu}-k_{a}k^{\mu}),
\end{equation}
implying that the object of anholonomity can also be described using only bilinears; via (\ref{Connection in local frame, final form}) the \emph{connection} in the local frame can also be described using only bilinears. Furthermore, the torsion field equation provides us with the constraints
\begin{align}
&\Omega_{a}=\frac{\rmi}{2}\Theta_{b}{}^{b}{}_{a}, \\
&\Omega_{5}^{d}=-\frac{1}{8}\Theta_{abc}\epsilon^{abcd}+\frac{3\pi}{2}Gk^{d}, \\
&\Gamma_{abc}=-\pi G\epsilon_{abcd}k^{d}+\frac{1}{2}(\Theta_{abc}-\Theta_{bca}-\Theta_{cab}).\label{Summary spin connection constraint}
\end{align}
Taken together, the equations (\ref{Summary Einstein's equations})-(\ref{Summary spin connection constraint}) describe the gravitationally self-interacting Einstein-Cartan-Dirac equations, in terms of the Lorentz covariant observables of the Dirac field: the Dirac bilinears. We believe the inverted forms of the Dirac equation (\ref{Summary inverted Dirac equation Omega}) and (\ref{Summary inverted Dirac equation Omega5}), the Fierz identities (\ref{Main Fierz identitity 1}) and (\ref{Main Fierz identitity 2}) that lead to their description in terms of Dirac bilinears as opposed to spinors, and their application to the Einstein-Cartan-Dirac system, to be new results.

In the electromagnetic case of the self-coupled Maxwell-Dirac equations, we showed that this system is able to be reduced in the presence of global spacetime symmetries corresponding to subgroups of the Poincar\'{e} group, and we gave four specific examples  \cite{InglisJarvis2014}. The approach we used was an infinitesimal method, which involved using the Lie generators of a particular Poincar\'{e} subalgebra, provided by Patera, Winternitz \& Zassenhaus \cite{PateraWinternitzZassenhaus1975}, to calculate joint invariant scalar and vector fields, which were then applied to the physical equations to obtain new exact and numerical solutions \cite{Inglis2015}. Due to the similar complexity of the Einstein-Cartan-Dirac equations, global symmetry reduction using the same techniques is one way in which solutions to this system can be pursued.

Another avenue of study which the results of this paper highlight is that of the extended Fierz algebra. The derivation of the Fierz identities needed to manipulate expressions involving Dirac bilinears, for the case where the spinors carry no tensor indices is straightforward (see (\ref{Fierz expansion})-(\ref{Bilinear trace identity})). However, the Dirac equation and related expressions of course involve partial derivatives of spinors, so that new classes of ``higher rank'' Fierz identities must be obtained. Equations (\ref{Main Fierz identitity 1}) and (\ref{Main Fierz identitity 2}) are two examples of a much broader set of such relations.

\section*{Acknowledgements}
The authors would like to thank Friedrich W. Hehl, for his valuable comments and advice in the development of this work.

\clearpage
\appendix
\section{\label{Glossary of symbols}Glossary of symbols}

\begin{table}[!h]
\caption{Glossary of symbols used in this article.}
\footnotesize{
\begin{tabular}{l l}
\vspace{-3mm}
\\
\hline
\vspace{-3mm}
\\
Symbol & Description \\
\vspace{-4mm}
\\
\hline
\vspace{-3.5mm}
\\
$\mu,\nu,...=0,1,2,3$&Global coordinate frame indices\\
$a,b,...=0,1,2,3$&Local orthonormal frame indices\\
$g_{\mu\nu}$&Global coordinate frame metric\\
$\eta_{ab}=diag(1,-1,-1,-1)$&Local orthonormal frame metric\\
$e_{a}{}^{\mu}$&Vierbein coordinate components\\
$\gamma^{a}$&Gamma matrices\\
$\gamma^{5}=\gamma_{5}=\rmi\gamma^{0}\gamma^{1}\gamma^{2}\gamma^{3}$&Gamma-5 matrix\\
$\sigma_{ab}=\frac{\rmi}{2}[\gamma_{a},\gamma_{b}]$&Rank-2 gamma matrix commutator\\
$\Gamma_{R=1,...,16}=\{I,\gamma^{a},\sigma^{ab},\gamma_{5}\gamma^{a},\gamma_{5}\}$&Dirac-Clifford algebra basis\\
$\epsilon^{abcd}$&Levi-Civita symbol\\
$\delta^{abcd}:=\rmi(\eta^{ac}\eta^{bd}-\eta^{ad}\eta^{bc})$&Sylvester tensor\\
$\Gamma_{\mu}=\frac{1}{2}\Gamma_{\mu}{}^{ab}S_{ab}$&Spinor connection\\
$S_{ab}=-\frac{\rmi}{2}\sigma_{ab}$&Lorentz generators in Dirac spinor rep.\\
$\Gamma_{\mu\nu}{}^{\lambda}$&Riemann-Cartan connection\\
$\psi,\chi$&Dirac spinors\\
$\psibar:=\psi^{\dagger}\gamma^{0}$&Dirac conjugate spinor\\
$\psi{}^{\rmc}:=C\psibar{}^{\rmT}=\rmi\gamma^{2}\gamma^{0}\psibar{}^{\rmT}$&Dirac charge conjugate spinor\\
$R^{\mu}{}_{\nu\sigma\lambda}$&Riemann-Cartan curvature tensor\\
$R_{\mu\nu}$&Ricci tensor\\
$R$&Ricci scalar\\
$\Lambda$&Cosmological constant\\
$\delta^{\mu}_{\nu}:=diag(1,1,1,1)$&Kronecker delta\\
$\Upsilon_{\mu\nu}{}^{\lambda}:=\Gamma^{\lambda}_{\mu\nu}-\Gamma^{\lambda}_{\nu\mu}$&Torsion tensor\\
$T^{\mu\nu}$&Stress-energy tensor\\
$\Sigma^{\mu\nu\lambda}$&Spin angular momentum density tensor\\
$\Omega_{d}:=\frac{1}{4}\delta_{adbc}\omega^{abc}$&Gravitational vector potential\\
$\Omega_{5d}:=\frac{1}{4}\epsilon_{abcd}\omega^{abc}$&Gravitational axial vector potential\\
$A_{a}$&Electromagnetic vector potential\\
$A_{5a}$&Abelian axial vector potential\\
$\sigma:=\psibar\psi$&Dirac bilinear scalar field\\
$\omega:=\psibar\gamma_{5}\psi$&Dirac bilinear pseudoscalar field\\
$j^{a}:=\psibar\gamma^{a}\psi$&Dirac bilinear vector field\\
$k^{a}:=\psibar\gamma_{5}\gamma^{a}\psi$&Dirac bilinear axial vector field\\
$s^{ab}:=\psibar\sigma^{ab}\psi$&Dirac bilinear rank-2 tensor field\\
$\sdual^{ab}:=\psibar\gamma_{5}\sigma^{ab}\psi$&Dirac bilinear dual rank-2 tensor field\\
$m^{a}:=\rmRe\{\psibar{}^{\rmc}\gamma^{a}\psi\}$&Complex Dirac bilinear, real part\\
$n^{a}:=\rmIm\{\psibar{}^{\rmc}\gamma^{a}\psi\}$&Complex Dirac bilinear, imaginary part\\
$t_{a}{}^{\mu}:=(\sigma^{2}-\omega^{2})^{-1/2}[j^{\mu},m^{\mu},n^{\mu},k^{\mu}]$&Dirac tetrad frame, coord. components\\
$\Theta_{abc}:=e_{a\mu}(\partial_{b}e_{c}{}^{\mu}-\partial_{c}e_{b}{}^{\mu})$&Object of anholonomity\\
$K_{abc}:=\frac{1}{2}(-\Upsilon_{abc}+\Upsilon_{bca}-\Upsilon_{cab})$&Contorsion tensor\\

\hline
\end{tabular}
}
\end{table}

\clearpage


\bibliographystyle{iopart-num}
\bibliography{Master_Bib_File}

\providecommand{\newblock}{}
\begin{thebibliography}{10}
\expandafter\ifx\csname url\endcsname\relax
  \def\url#1{{\tt #1}}\fi
\expandafter\ifx\csname urlprefix\endcsname\relax\def\urlprefix{URL }\fi
\providecommand{\eprint}[2][]{\url{#2}}

\bibitem{LandauLifshitz1971}
Landau L~D, Pitaevskii L~P, Lifshitz E~M and Berestetskii V~B 1971 {\em
  Relativistic quantum theory\/} (Pergamon)

\bibitem{Bethe1947}
Bethe H~A 1947 {\em Phys. Rev.\/} {\bf 72} 339--41

\bibitem{Eliezer1958}
Eliezer C~J 1958 {\em Proc. Camb. Phil. Soc.\/} {\bf 54} 247

\bibitem{BoothLeggJarvis2001}
Booth H~S, Legg G and Jarvis P~D 2001 {\em J. Phys. A: Math. Gen.\/} {\bf 34}
  5667

\bibitem{Takabayasi1957}
Takabayasi T 1957 {\em Prog. Theor. Phys. Supplement\/} {\bf 4} 1

\bibitem{Fierz1937}
Fierz M 1937 {\em Z. Phys.\/} {\bf 104} 553

\bibitem{Lounesto2001}
Lounesto P 2001 {\em Clifford algebras and spinors\/} (Cambridge university
  press)

\bibitem{Baylis1996}
Baylis W~E 1996 {\em Clifford (geometric) algebras: with applications in
  physics, mathematics, and engineering\/} (Birkh\"{a}user)

\bibitem{Crawford1985}
Crawford J~P 1985 {\em J. Math. Phys.\/} {\bf 26} 1439

\bibitem{Takahashi1983}
Takahashi Y 1983 {\em J. Math. Phys.\/} {\bf 24} 1783

\bibitem{InglisJarvis2014}
Inglis S~M and Jarvis P~D 2014 {\em Ann. Phys.\/} {\bf 348} 176--222

\bibitem{PateraWinternitzZassenhaus1975}
Patera J, Winternitz P and Zassenhaus H 1975 {\em J. Math. Phys.\/} {\bf 16}
  1597--1614

\bibitem{InglisJarvis2012}
Inglis S~M and Jarvis P~D 2012 {\em J. Phys. A: Math. Theor.\/} {\bf 45} 465202

\bibitem{BistrovicJackiwNairPi2003}
Bistrovic B, Jackiw R, Li H, Nair V~P and Pi S~Y 2003 {\em Phys. Rev. D\/} {\bf
  67} 025013

\bibitem{BlagojevicHehl2013}
Blagojevi\'{c} M and Hehl F~W 2013 {\em Gauge theories of gravitation: a reader
  with commentaries\/} (World Scientific, Imperial College Press, London)

\bibitem{Goedecke1974}
Goedecke G~H 1974 {\em J. Math. Phys.\/} {\bf 15} 792

\bibitem{HehlHeydeKerlickNester1976}
Hehl F~W, Heyde P~V~D, Kerlick G~D and Nester J~M 1976 {\em Rev. Mod. Phys.\/}
  {\bf 48} 393--416

\bibitem{Crawford1993}
Crawford J~P 1993 Local automorphism invariance: a generalization of general
  relativity {\em Clifford algebras and their applications in mathematical
  physics\/} (Springer) pp 261--68

\bibitem{Yepez2011}
Yepez J 2011 {\em arXiv:1106.2037\/}

\bibitem{HehlLemkeMielke1991}
Hehl F~W, Lemke J and Mielke E~W 1991 Two lectures on fermions and gravity {\em
  Geometry and theoretical physics\/} (Springer) pp 56--140

\bibitem{Kibble1961}
Kibble T~W~B 1961 {\em J. Math. Phys.\/} {\bf 2} 212--21

\bibitem{ItzyksonZuber1980}
Itzykson C and Zuber J~B 1980 {\em Quantum Field Theory\/} (McGraw-Hill
  International Book Co. New York)

\bibitem{Zund1976}
Zund J~D 1976 {\em Ann. Mat. Pura Appl.\/} {\bf 110} 29--135

\bibitem{HehlDatta1971}
Hehl F~W and Datta B~K 1971 {\em J. Math. Phys.\/} {\bf 12} 1334--39

\bibitem{Pollock2010}
Pollock M~D 2010 {\em Acta Phys. Polon.\/} {\bf 41} 1827--46

\bibitem{Zhelnorovich1965}
Zhelnorovich V~A 1966 {\em J. App. Math. Mech.\/} {\bf 30} 1289--1300

\bibitem{Halbwachs1960}
Halbwachs F 1960 {\em Th{\'e}orie relativiste des fluides {\`a} spin\/}
  (Gauthier-Villars)

\bibitem{InglisJarvis2016}
Inglis S~M and Jarvis P~D 2016 {\em Ann. Phys.\/} {\bf 366} 57--75

\bibitem{AliCafaroCapozzielloCorda2009}
Ali S~A, Cafaro C, Capozziello S and Corda C 2009 {\em Int. J. Theor. Phys.\/}
  {\bf 48} 3426--48

\bibitem{Blagojevic2003}
Blagojevi\'{c} M 2003 {\em arXiv:gr-qc/0302040\/}

\bibitem{Kaempffer1981}
Kaempffer F~A 1981 {\em Phys. Rev. D\/} {\bf 23} 918

\bibitem{Inglis2015}
Inglis S~M 2015 {\em The manifestly gauge invariant {M}axwell-{D}irac
  equations\/} (Ph.D. Thesis: University of Tasmania)

\end{thebibliography}

\end{document}